\documentclass[5p, twocolumn]{elsarticle}
\usepackage{lineno}

\usepackage{graphicx}

\usepackage{dcolumn}

\usepackage{bm}

\usepackage{braket}
\usepackage{color}
\usepackage{ulem}
\usepackage{comment}
\usepackage{amsmath}
\usepackage[final]{showlabels}
\usepackage{mathtools}

\begin{document}
\begin{frontmatter}

\title{Measurement of nuclear spin relaxation time in lanthanum aluminate for development of polarized lanthanum target}

\author[nagoya]{K.~Ishizaki\corref{cor}}
\ead{ishizaki@phi.phys.nagoya-u.ac.jp}
\cortext[cor]{Corresponding author}

\address[nagoya]{Nagoya University, Furocho, Chikusa, Nagoya, 464-8602, Japan}
\address[hiroshima]{Hiroshima University, Kagamiyama, Higashi-hiroshima, 739-8527, Japan}
\address[yamagata]{Yamagata University, Koshirakawa, Yamagata 990-8560, Japan}
\address[KMI]{KMI, Nagoya University, Furocho, Chikusa, Nagoya, 464-8602, Japan}
\address[RCNP]{Research Center for Nuclear Physics, Osaka University, Ibaraki, Osaka, 567-0047, Japan}
\address[UKHospital]{The University of Tokyo Hospital, Hongo, Bunkyo, Tokyo, 113-8655, Japan}

\author[nagoya]{H.~Hotta}%
\author[nagoya]{I.~Ide}%
\author[hiroshima]{M.~Iinuma}%
\author[yamagata]{T.~Iwata}%
\author[KMI]{M.~Kitaguchi}%
\author[RCNP,nagoya]{H.~Kohri}%
\author[yamagata]{D.~Miura}%
\author[yamagata]{Y.~Miyachi}%
\author[UKHospital]{T.~Ohta}%
\author[nagoya]{H.~M.~Shimizu}%
\author[RCNP]{H.~Yoshikawa}%
\author[RCNP]{M.~Yosoi}%

\date{\today}

\begin{abstract}
The nuclear spin-lattice relaxation time ($T_1$) of lanthanum and aluminum nuclei in a single crystal of lanthanum aluminate doped with neodymium ions is studied to estimate the feasibility of the dynamically polarized lanthanum target applicable to beam experiments.
The application of our interest is the study of fundamental discrete symmetries in the spin optics of epithermal neutrons. This study requires a highly flexible choice of the applied magnetic field for neutron spin control and favors longer $T_1$ under lower magnetic field and at higher temperature.
The $T_1$ of $^{139}{\rm La}$ and ${}^{27}{\rm Al}$ was measured under magnetic fields of $0.5$-$2.5$ T and at temperatures of $0.1$-$1.5$ K and found widely distributed up to 100 h.
The result suggests that the $T_1$ can be as long as $T_1 \sim$ 1 h at $0.1$ K with a magnetic field of $0.1$ T, which partially fulfills the requirement of the neutron beam experiment.
Possible improvements to achieve a longer $T_1$ are discussed.

\end{abstract}



\end{frontmatter}



\section{Introduction}

The feasibility of the polarized nuclear target of ${}^{139}{\rm La}$ is explored for the study of the spin-related correlation terms in compound nuclear states induced by polarized epithermal neutrons, which introduces an enhanced sensitivity to the breaking of spatial and time-reversal symmetries.
Enhanced parity-nonconservation (PNC) effects are observed in the compound resonances of medium heavy nuclei and the enhancement factor reaches $10^6$ compared with the PNC effect in the nucleon-nucleon interaction~\cite{Alfimenkov,mit01}.
The enhancement mechanism is further studied for ${}^{139}{\rm La}$ by observing the neutron spin dependence and angular distribution of $\gamma$-rays in the reaction ${}^{139}{\rm La}({\rm n},\gamma){}^{140}{\rm La}$. These studies are conducted to quantitatively estimate the possible enhancement of the time-reversal violation (T-violation) in the optical behavior of the neutron spin on transmission through a polarized target~\cite{Gudkov, BL04, Yamamoto}.

To polarize the nuclear spins of $^{139}$La, two well-known techniques, static nuclear polarization (SNP), and dynamic nuclear polarization (DNP), are applicable. In SNP, preparing an environment of a huge magnetic field and an ultra-low temperature makes it possible to obtain high nuclear polarization in thermal equilibrium.
In the case of a hydrogen deuteride (HD) target~\cite{Hoblit,Ho,Kohri}, for example, the nuclear relaxation time is approximately a few hours before the aging of the HD target and longer than several months after aging. This is a special case because, generally, SNP requires a long aging time to achieve high thermal polarization. In the case of $^{139}$La, a magnetic field of 17 T and temperature of 0.01 K can provide a polarization of approximately 59\%.
One candidate target for the beam experiments is a metal target, which has been used in some beam experiments~\cite{Alfimenkov_PT, Roubean, Glattli}, because metal targets have a considerably shorter relaxation time than the insulating materials. However, keeping high polarization under a low external magnetic field is difficult during the beam experiments. 

The dynamic nuclear polarization (DNP) is a method to dynamically polarize nuclei by transferring the polarization of the paramagnetic centers to the nuclear polarization via a double spin-flip process induced by microwave irradiation, and it is appropriate to fulfill the requirements of the T-violation experiment.
Lanthanum aluminate doped with neodymium ions La$_{1-x}$Nd$^{3+}_{x}$AlO$_3$ has been studied as the candidate material for the polarized lanthanum target after the dynamical enhancement of $^{139}$La and $^{27}$Al was successfully observed at the Paul Scherrer Institute (PSI).
The achieved polarization was approximately 50\% in the $^{139}$La spin ($I$ = 7/2) with the dopant content of 0.03 mol\% $(x=3\times 10^{-4})$ at a magnetic field of 2.35 T and a temperature lower than 0.3 K~\cite{Hautle}.
The LaAlO$_3$ has an additional advantage that the $^{139}$La in LaAlO$_3$ can avoid the decrease of the vector polarization, which is caused by the mixing of the eigenstates of the nuclear spin by the nuclear quadrupole interaction with the electric field gradient~\cite{Takahashi1993}.
Here, we mention that the La$_2$Mg$_3$(NO$_3$)$_{12}\cdot$ 24H$_2$O is not an appropriate compound because its lanthanum content is small, although it is a well known lanthanum compound to which the DNP can be applied.

The nuclear spin-lattice relaxation time ($T_1$) of $^{139}$La is the remaining key parameter to be studied for the estimation of the feasibility and applicability of LaAlO$_3$ to the T-violation experiment.
It was measured above 1 K~\cite{Maekawa1995}, but additional measurement below 1 K is necessary for practical installation to the beam experiment, which requires a long-term stability to maintain the ${}^{139}{\rm La}$ polarization with the magnetic field as low as 0.1 T required for the neutron spin control.

Generally, the measurements of $T_{1}$ need sufficiently large NMR signals. Because the thermal NMR signals of $^{139}$La 
are likely weak because of the existence of eight sublevels, the $T_{1}$ measurements may be difficult depending on the 
measurement conditions. For the LaAlO$_{3}$ crystal, however, studies on the $T_{1}$ of $^{27}$Al, 
which is easily measurable, help in the evaluation of the $T_1$ of $^{139}$La under the target condition 
because the nuclear relaxation usually originates from the paramagnetic dopant. 

In this paper, we report the temperature and magnetic-field dependence of $T_1$ of $^{139}$La and $^{27}$Al in the Nd-doped LaAlO$_3$ crystal over a temperature range of 0.1--1.5 K and magnetic fields from 0.5 to 2.5 T.  
We estimate the $T_1$ of $^{139}$La under low magnetic fields that are desirable in T-violation experiments.

\section{Experiment}
\subsection{Cryostat}

We used two types of refrigerators.
One is a $^{3}$He-$^{4}$He dilution refrigerator (DRS) manufactured by Leiden Cryogenics B.V.~\cite{Leiden}. This refrigerator covers the temperature range above 6 mK and is equipped with a superconducting solenoid (NbTi and Nb$_{3}$Sn) produced by JASTEC Co., Ltd.~\cite{Jastec} in Japan. This magnet can generate a maximum magnetic field of 17 T. 
The other is a $^{4}$He pumping cryostat, which is referred to as a storage cryostat (SC), produced by Oxford Instruments~\cite{Oxford}. 
This cryostat can provide an environment at a temperature of 1.5 K under magnetic fields up to 2.5 T. The SC is normally used 
for transporting polarized HD targets from Osaka University to SPring-8~\cite{Kohri}. 

At 0.1 K and 0.5 K,
we excited the superconducting solenoid of the DRS and performed nuclear magnetic resonance (NMR) measurements with a sweep of the magnetic field.
The temperature of the mixing chamber was stabilized by switching a heater mounted there. To determine the sample temperature, 
we used the readout from a carbon resistance thermometer, which was attached to the mixing chamber.    
In the field sweep, the induced Eddy current generally leads to an uncertainty in temperature. 
The uncertainty was approximately 0.01 K in the experiments. 

\begin{figure}[h]
  \centering
  \includegraphics[width=7.0cm]{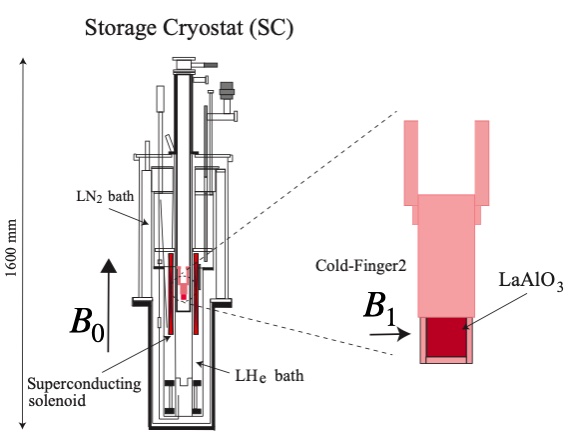}
  \caption{LaAlO$_{3}$ crystal installed at the bottom of the 
           storage cryostat (SC). 
           The height of the SC is approximately 1.6 m, and the bottom part 
           has the lowest temperature of 1.5 K. }
  \label{fig:scfigure} 
\end{figure}

The NMR measurement was also performed in the SC refrigerator using the same method as that in the DRS. 
The uncertainty of the SC temperature was less than approximately 0.05 K. The temperature of the sample 
was monitored by reading a RuO$_2$ resistance thermometer attached to the Cold-finger2, as shown in Fig.~\ref{fig:scfigure}.

\subsection{NMR Systems}

We used a portable NMR system~\cite{Ohta1,Ohta2}, which had been developed
for a polarized HD target~\cite{Kohri,Yanai}, to measure
the relaxation times of $^{139}$La and $^{27}$Al. 

In the NMR measurements with the DRS,
two coaxial cables were used to form a connection between a pickup coil and a tuning circuit placed outside the cryostat. 
One was a semi-flexible coaxial cable (MULTI-FLEX 141 produced by Suhner) with a length of 2 m and a diameter of 4.2 mm. 
This cable was used in the range of 293 -- 301 K. 
The capacitance of the cable was 95.0 pF/m.
The other was a semi-rigid coaxial cable (SC-119/50-SCN-CN produced by Coax Co., Ltd.) with a length of 2.5 m 
and a diameter of 1.19 mm. The semi-rigid cable was used inside the cryostat.
The capacitance of the cable is 95.2 pF/m.
The measured self-inductance of the pickup coil holding the LaAlO$_3$ crystal was 5.40 $\mu$H.
In the measurements with the SC, a semi-rigid coaxial cable (EZ47-AL-TP produced by EZ Form Cable Corporation) with a length of 1 m and 
a diameter of 1.19 mm was used instead of the SC-119/50-SCN-CN coaxial cable. 

In all the NMR measurements, the strength of the applied magnetic field was measured by reading the current supplied to the superconducting magnet. 
The uncertainty of the magnetic field strength was approximately 1\%.
The continuous-wave radio frequency (CW-RF) applied through the pickup coil was appropriately attenuated to avoid affecting the $T_{1}$ measurements. 

\subsection{Sample}

The sample is a LaAlO$_3$ crystal with Nd-dopant content of 0.03 mol\% ($x=3\times10^{-4}$) with dimensions of 1.5 $\times$ 1.5 $\times$ 1.5 cm$^3$ and features twin domains~\cite{shinkosha}. The Nd content was analyzed using inductively coupled plasma mass spectrometry (ICP-MS)~\cite{Clearize}. 
The results confirmed that the amount of paramagnetic impurities including Ir, W, and Mo, which are typical materials used to build the crucible, is much less than the amount of Nd. The sample was used with the twin structure still contained.

\begin{figure}[h]
  \centering
  \includegraphics[width=7.0cm]{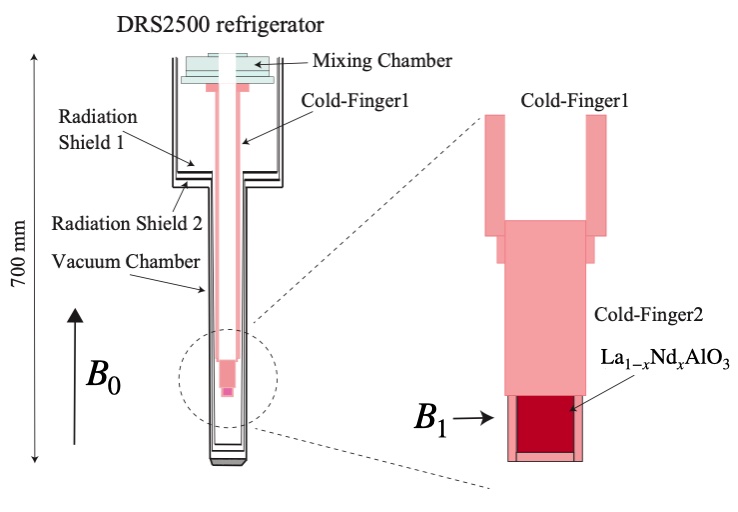}
  \caption{LaAlO$_{3}$ crystal installed at the bottom of 
           the DRS. 
           The height of the DRS refrigerator is approximately 2.5 m, and 
           the mixing chamber, where $^{3}$He and $^{4}$He are mixed, 
           has the lowest temperature. }
  \label{fig:crystal} 
\end{figure}

\begin{figure}[h]
  \centering
  \includegraphics[width=7.0cm]{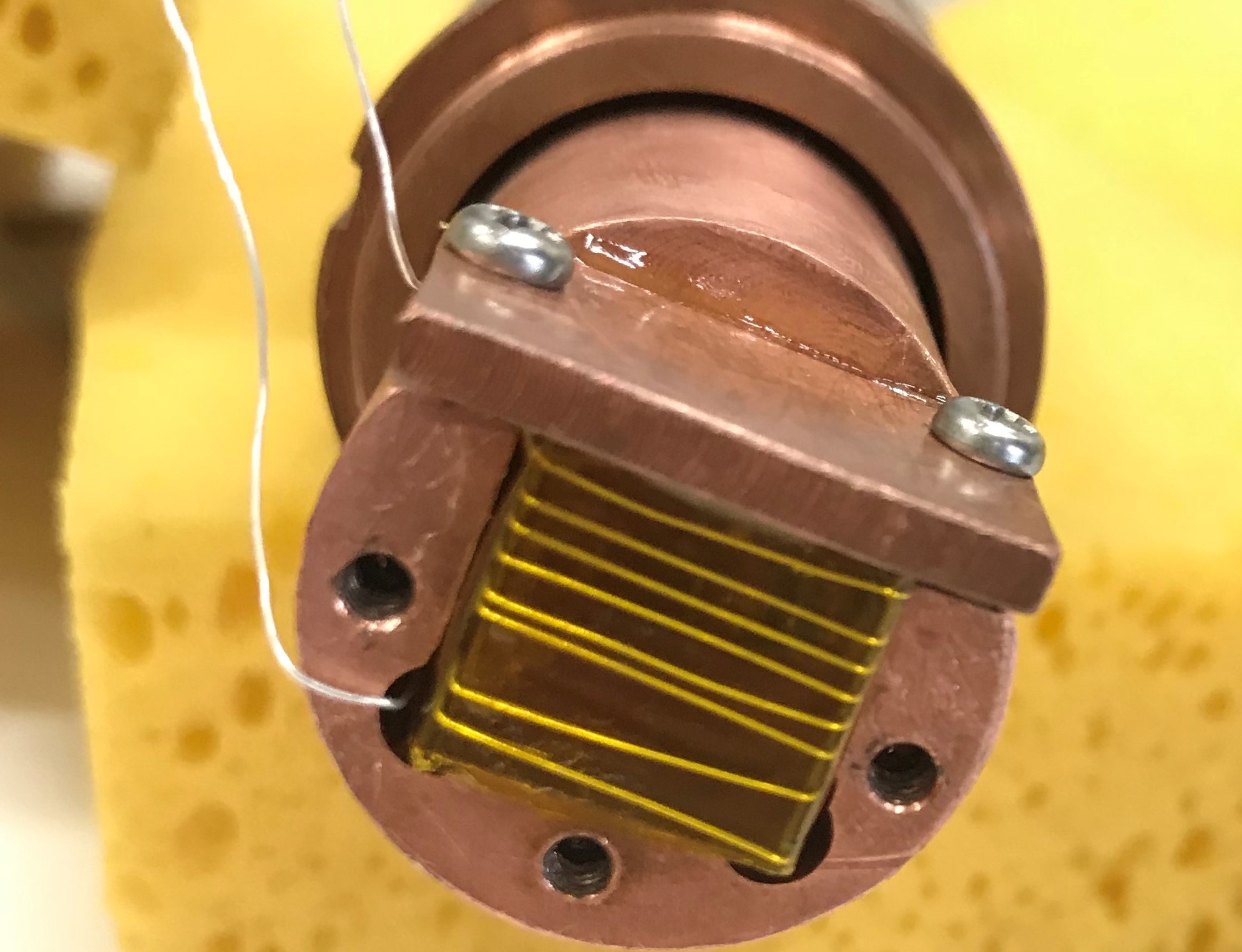}
  \caption{LaAlO$_3$ crystal attached to Cold-finger2 and the pickup coil. In the experiment, the crystal was covered by a thin plate, which was attached to the kapton surface at the front of this figure to fold the sample.}
  \label{fig:sample} 
\end{figure}

Cold-Finger2 was attached to Cold-Finger1, which was thermally connected to the mixing chamber, to install the LaAlO$_{3}$ crystal at the bottom of the DRS, as shown in Fig.~\ref{fig:crystal}. 
The crystal was wound with eight turns of a Teflon-coated silver wire 0.3 mm in diameter to detect the NMR signal, as shown in Fig.~\ref{fig:sample}. The wire was fixed with a kapton tape to maintain the shape of the pickup coil. The surfaces of the sample were kept in contact with the surface of the Cold-Finger2 using Apiezon-N grease to increase the cooling efficiency. 

\subsection{Procedure}

To obtain the $T_{1}$ from the time dependence of many NMR spectra, we should evaluate the height and width of each peak as accurately as possible. Each peak of $^{139}$La was individually analyzed by fitting with a linear combination of Lorentzian and linear functions as follows:

\begin{equation}
  \begin{split}
      f_i(B) = I_i \frac{w^2}{(B-B_i)^2+w^2} + a (B-B_i) + b, 
    \label{eq:fit_func_La}
  \end{split}
\end{equation}
where $B_{i}$ is the centroid of each signal peak, 
$I_{i}$ is the height of the signal peak, 
$w$ is a parameter related to the peak width, and 
$a$ and $b$ are parameters of the linear function for evaluating the background.

In the $^{27}$Al NMR measurements, the peaks are not completely separated, because the homogeneity of the magnetic field was insufficient. In this case, a combination of multi Gaussian and linear functions is used as the fitting function, which can be expressed as follows:

\begin{equation}
\begin{split}
    f(B) = \sum_i I_i \exp \left( -\frac{(B - B_i)^2}{2\sigma^2} \right) + a (B-B_5) + b, 
  \label{eq:fit_func_Al}
\end{split}
\end{equation}
where $B_{i}$ ($B_{1}<B_{2}<\cdot\cdot\cdot<B_{9}$) is the centroid of each signal peak, 
$I_{i}$ is the height of the signal peak, 
$\sigma$ is the standard deviation, and 
$a$ and $b$ are parameters of the linear function for evaluating the background. The Lorentzian fitting for the $^{27}$Al NMR spectrum was less accurate than in the Gaussian case.

The relaxation rate $\Gamma \, (=1/T_{1})$ is obtained by fitting the following simple exponential function to the time dependence of the evaluated intensities:

\begin{equation}
  I(t) = (I_0 - I_{\rm eq}) \exp (-\Gamma t) + I_{\rm eq}, 
  \label{eq:time_evolution}
\end{equation}
where $I_0$ and $I_{\rm eq}$ are the evaluated intensity of the initial polarization and that at thermal equilibrium, respectively. 

\begin{figure}[h]
  \centering
  \includegraphics[width=7.0cm]{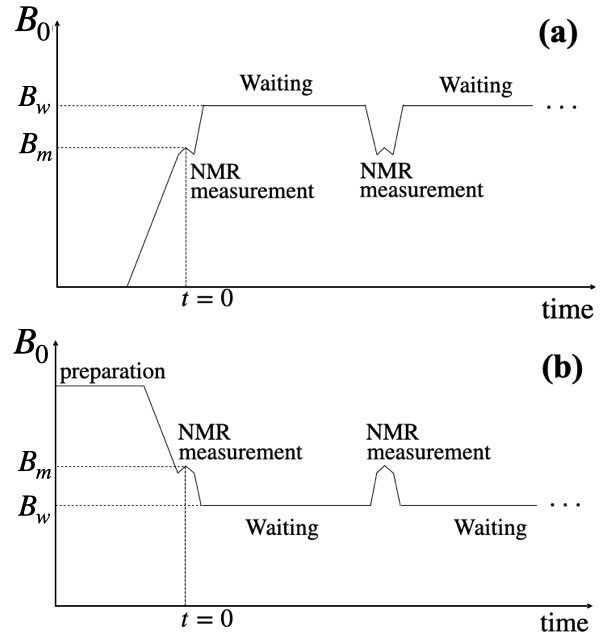}
  \caption{Sequence of the $T_1$ measurement with CW-NMR for (a) the "buildup" method and (b) the "decay" method. The NMR spectra were obtained when $B_0$ was equal to $B_{\rm m}$. While the "Waiting", we waited for the change of the nuclear polarization at $B_{\rm w}$. Thus, the measurements in the sequence provided the $T_1$ at $B_{\rm w}$.
  }
  \label{fig:measurement_sequence} 
\end{figure}

The measurements were performed as follows.
\begin{enumerate}
\item The sample is cooled down to a temperature suitable for the $T_1$ measurements. 
\item To prepare the initial polarization, the crystal is aged for more than half a day under a magnetic field higher than that at the thermal equilibrium or without a magnetic field. 
\item The NMR circuit is tuned to a frequency selected for the NMR detection. 
\item The magnetic field is changed to the vicinity of the field corresponding to the 
      $^{139}$La or $^{27}$Al resonance. 
\item Some NMR data are acquired by sweeping the field at a speed of 0.034 T/min. 
\item The magnetic field is changed to the condition for the $T_{1}$ measurement. Subsequently, 
      the crystal is left for some time.
\item Steps 4 to 6 are repeated for a long term.  
\end{enumerate}
We measured the buildup and decay toward thermal equilibrium of the NMR signals as shown in Fig.~\ref{fig:measurement_sequence}.

We measured the $T_1$ of $^{139}$La at four conditions and that of $^{27}$Al at seven conditions, as presented in Table~\ref{table:dataset}.

\begin{table}[h]
  \centering
    \caption{Summary of the data set. The terms "buildup" and "decay" indicate the method of $T_1$ measurements. $f$ is the RF frequency of the NMR measurement.}
    \begin{tabular}{ccccc} \hline
    \textrm{Nuclei}&
    \textrm{$T$ [K]}&
    \textrm{$B$ [T]}&
    \textrm{$f$ [MHz]}&
    \textrm{Method}\\
    \hline
    $^{139}\mathrm{La}$ & 0.5 & 2.5 & 16.1 & decay\\
                       & 0.5 & 1.0 & 5.7 & decay\\
                       & 0.5 & 0.5 & 5.6 & decay\\
                       & 0.1 & 0.75 & 5.6 & buildup\\ 
     $^{27}\mathrm{Al}$ & 1.5 & 2.1 & 23.8 & buildup\\ 
                       & 1.5 & 1.0 & 11.0 & decay\\
                       & 0.5 & 2.5 & 28.2 & decay\\
                       & 0.5 & 1.0 & 11.8 & decay\\
                       & 0.5 & 0.5 & 9.1 & decay\\
                       & 0.1 & 1.6 & 16.8 & buildup\\
                       & 0.1 & 0.76 & 16.4 & buildup\\
    \hline
    \end{tabular}
    \label{table:dataset}
\end{table}

\section{Results}
\label{sec:result}

\subsection{NMR spectra}

\begin{figure*}[h]
  \centering
  \includegraphics[width=14.0cm]{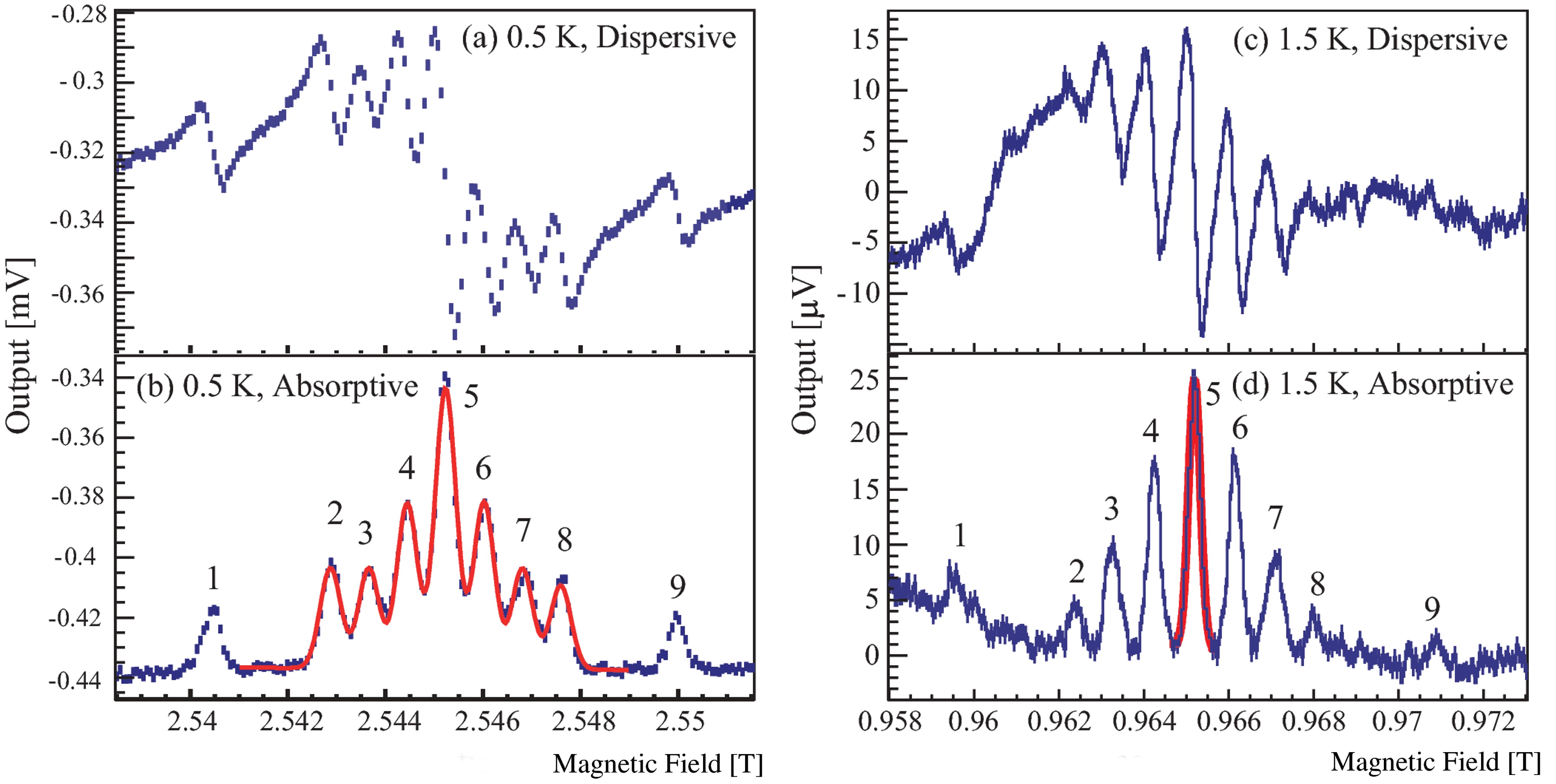}
  \caption{Typical $^{27}$Al NMR spectra. The spectra (a) and (b) were obtained in the DRS at 0.5 K. The spectra (c) and (d) were obtained in the SC at 1.5 K.}
  \label{fig:alnmr} 
\end{figure*}

Fig.~\ref{fig:alnmr} shows the dispersive and absorptive NMR spectra for $^{27}$Al($I=5/2$) at 0.5 K with the DRS and at 1.5 K with the SC. 
Nine peaks are observed.
On the absorptive spectrum of seven central peaks, Eq.~(\ref{eq:fit_func_Al}) was used as the fitting function. 
To reduce the complexity of the fitting, we assumed that the widths of all the peaks and the spacings between the neighboring peaks were identical. 
In contrast, peak 1 and peak 9, we individually fitted a linear combination of Gaussian and linear functions.

Consequently, the peak width is obtained as $\sigma=(2.12 \pm 0.01) \times 10^{-4}$ T, which is almost consistent with the field homogeneity of approximately 10$^{-4}$ in the DRS. 
The distances between neighboring peaks among the seven peaks were obtained as 
(7.86$\pm$0.01)$\times$10$^{-4}$ T. The distances between peak 1 and peak 2 and between peak 8 and peak 9 were 
(2.42$\pm$0.01)$\times$10$^{-3}$ T
and (2.38$\pm$0.01)$\times$10$^{-3}$ T, respectively, 
which were approximately three times 
larger than those between the neighboring peaks.
The $\sigma$ of the seven peaks and those of peak 1 and peak 9 were almost equal.
From similar analyses of the NMR spectra, we confirmed that all widths are independent of the elapsed time.

\begin{figure*}[h]
  \centering
  \includegraphics[width=14.0cm]{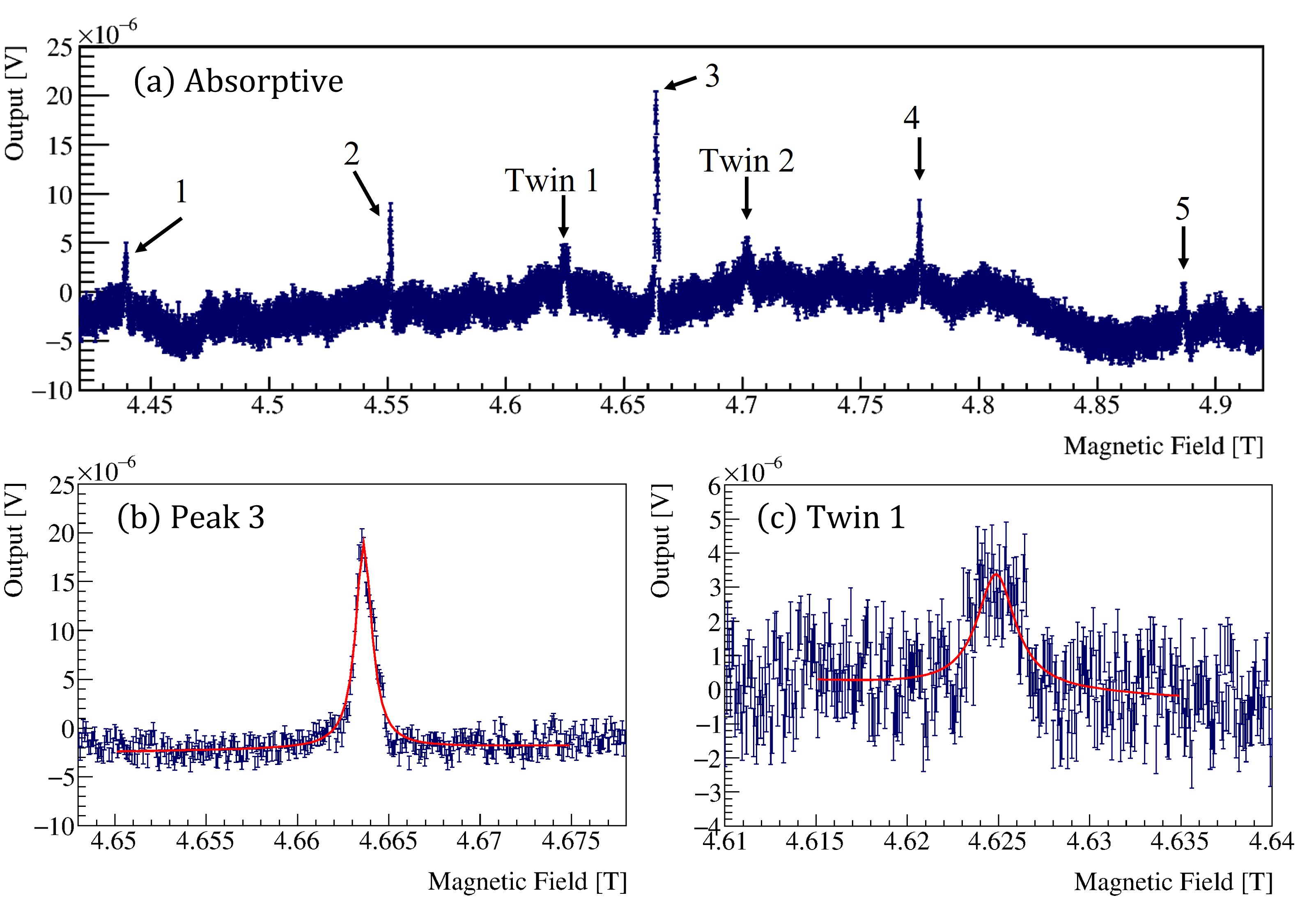}
  \caption{$^{139}$La NMR spectrum at approximately 5 T and 0.5 K. (a) : Absorptive part of $^{139}$La NMR. (b) : Fit to "Peak 3" that corresponds to the transition $-1/2 \leftrightarrow 1/2$. (c) : Fit to "Twin 1".
  }
  \label{fig:lanmr} 
\end{figure*}


The (c) and (d) spectra in Fig.~\ref{fig:alnmr} were obtained in the SC at 1.5 K.
Because the superconducting solenoid of the SC has better homogeneity than that of the DRS2500 refrigerator,  
the peaks are well separated from the neighboring peaks. 
Only the central peak is fitted by a Gaussian with a linear function.

For the peak identification in $^{139}$La($I=7/2$), we measured broad NMR spectra at 5 T at 0.5 K with high sensitivity separately from the systematic $T_1$ measurements. Fig.~\ref{fig:lanmr} clearly shows five peaks, indicated by 1, 2, 3, 4, and 5, are separated by the same distance. 


The LaAlO$_{3}$ crystal has a distorted perovskite structure with a three fold rotation axis ($C_3$ axis) at temperatures lower than 813 K~\cite{Hayward}. 
Both the La and Al sites have an electric-field gradient with uniaxial symmetry. 
If a nucleus with a spin $I$ and a substate $m$ is in the electric-field gradient 
with the external magnetic field $B_{0}$, the resonance frequency of the transitions 
m+1$\leftrightarrow$m is 

\begin{equation}
  \nu_{m+1 \leftrightarrow m} = \nu_L - \nu_Q \left( \frac{3\cos ^2 \theta -1}{2} \right) \left( m + \frac{1}{2} \right), 
  \label{eq:resonance_frequency}
\end{equation}
where $\nu_{L}$ = $\gamma B_{0}/2\pi$ is the Larmor frequency of the nucleus,  
$\nu_{Q}$ is the quadrupole splitting frequency, and 
$\theta$ is the angle between the symmetrical axis of the electric-field gradient and direction of $B_0$~\cite{Hautle}.  

Generally, single crystals of LaAlO$_3$ naturally contain twin domain structures, which are commonly referred to as twinning domains. These domains have a symmetrical axis of the
inner electric-field gradient, and their orientations are different from each other. The angle between the two axes 
is calculated to be approximately $\arccos(1/3)$ from the structural phase transition of the crystal, 
which naturally occurs in the cooling process immediately after the crystal growth. As a result, the symmetric axis in one domain is parallel to the direction of $B_0$, whereas the other domain's axis is at an angle of $\arccos(1/3)$ from the direction of $B_0$. To distinguish them, henceforth, the former is referred to as the "primary domain" and the latter is the "secondary domain" in this paper. 

It becomes possible to identify the observed peaks in Fig.~\ref{fig:alnmr} and Fig.~\ref{fig:lanmr} by comparing the parameters on the peaks to the characteristics listed in Tables~\ref{table:peakcharacteristic} and \ref{table:la-peakcharacteristic}. In the $^{139}$La spectra, 
Table~\ref{table:la-peakcharacteristic} lists the value of the peak parameters, which are useful for identifying the peaks listed in Table~\ref{table:lapeaks}.
From the consistency of the values of $B_i-B_3$ with Table~\ref{table:la-peakcharacteristic}, the seven observed peaks are successfully identified as indicated in Fig.~\ref{fig:lanmr}. 
The other six peaks of $^{139}$La are not observed in the spectrum, because of the weakness of the NMR signals.

\begin{table}[h]
  \centering
  \caption{
  Characteristics of nine peaks that correspond to the visible peaks in actual $^{27}$Al NMR spectra. 
  The $B_i-B_5$ values were calculated from Eq.(\ref{eq:resonance_frequency}) 
  and NMR parameters in Ref.~\cite{Hautle}. 
  (p) and (s) are the primary domain and secondary domain, respectively.
  See text for more details.
  }
  \begin{tabular}{cccc}
  \hline
  \textrm{
    \begin{tabular}{c}
      Peak \\ index $i$
    \end{tabular}
  }&
  \textrm{Transition}&
  \textrm{Domain}&
  \textrm{$B_i-B_5$ [mT]}\\
  \hline
  1 & $-5/2 \leftrightarrow -3/2$ & (p) & -5.4\\
  2 & $-3/2 \leftrightarrow -1/2$ & (p) & -2.7\\
  3 & $-5/2 \leftrightarrow -3/2$ & (s) & -1.8\\
  4 & $-3/2 \leftrightarrow -1/2$ & (s) & -0.9\\
  5 & $-1/2 \leftrightarrow +1/2$  & (p), (s) & 0\\
  6 & $+1/2 \leftrightarrow +3/2$ & (s) & +0.9\\
  7 & $+3/2 \leftrightarrow +5/2$ & (s) & +1.8\\
  8 & $+1/2 \leftrightarrow +3/2$ & (p) & +2.7\\
  9 & $+3/2 \leftrightarrow +5/2$ & (p) & +5.4\\ \hline
  \end{tabular}
  \label{table:peakcharacteristic}
\end{table}

\begin{table}[h]
  \centering
  \caption{
    Characteristics of seven peaks among thirteen ones that correspond to the visible peaks in actual $^{139}$La NMR spectra. 
  (p) and (s) indicate the primary domain and secondary domain, respectively.
  The $B_i-B_3$ values were calculated from Eq.(\ref{eq:resonance_frequency}) 
  and NMR parameters in Ref.~\cite{Hautle}. 
  }
  \begin{tabular}{cccc}
  \hline
  \textrm{
    \begin{tabular}{c}
      Peak \\ index $i$
    \end{tabular}
  }&
  \textrm{Transition}&
  \textrm{Domain}&
  \textrm{$B_i-B_3$ [T]}\\
  \hline
  1 & $-5/2 \leftrightarrow -3/2$ & (p) & -0.224\\
  2 & $-3/2 \leftrightarrow -1/2$ & (p) & -0.112\\
  twin~1 & $-3/2 \leftrightarrow -1/2$ & (s) & -0.037\\
  3 & $-1/2 \leftrightarrow +1/2$  & (p), (s) & 0\\
  twin~2 & $+3/2 \leftrightarrow +5/2$ & (s) & +0.037\\
  4 & $+1/2 \leftrightarrow +3/2$ & (p) & +0.112\\
  5 & $+3/2 \leftrightarrow +5/2$ & (p) & +0.224\\ \hline
  \end{tabular}
  \label{table:la-peakcharacteristic}
\end{table}

\begin{table}[h]
  \centering
    \caption{Characteristics of seven peaks observed in the NMR measurement of $^{139}$La.}
    \begin{tabular}{cccc} \hline
    \textrm{Peak index $i$}&
    \textrm{$B_i$ [T]}&
    \textrm{$B_i -B_3$ [T]}&
    \textrm{Width [mT]}\\
    \hline
    1 & 4.4505 & -0.217 & $1.1 \pm 0.3$ \\
    2 & 4.5588 & -0.109 & $0.9 \pm 0.0$\\
    twin 1 & 4.6320 & -0.036 & $1.5 \pm 0.0 $\\
    3 & 4.6677 & 0 & $1.0 \pm 0.0$\\
    twin2 & - & - & - \\
    4 & 4.7774 & 0.110 & $1.1 \pm 0.1$ \\
    5 & 4.8879 & 0.220 & $1.1 \pm 0.1$ \\
    \hline
    \end{tabular}
    \label{table:lapeaks}
\end{table}

\subsection{$T_1$ measurements of $^{139}$La NMR in the DRS2500 refrigerator}
\begin{figure*}[h]
  \centering
  \includegraphics[width=14.0cm]{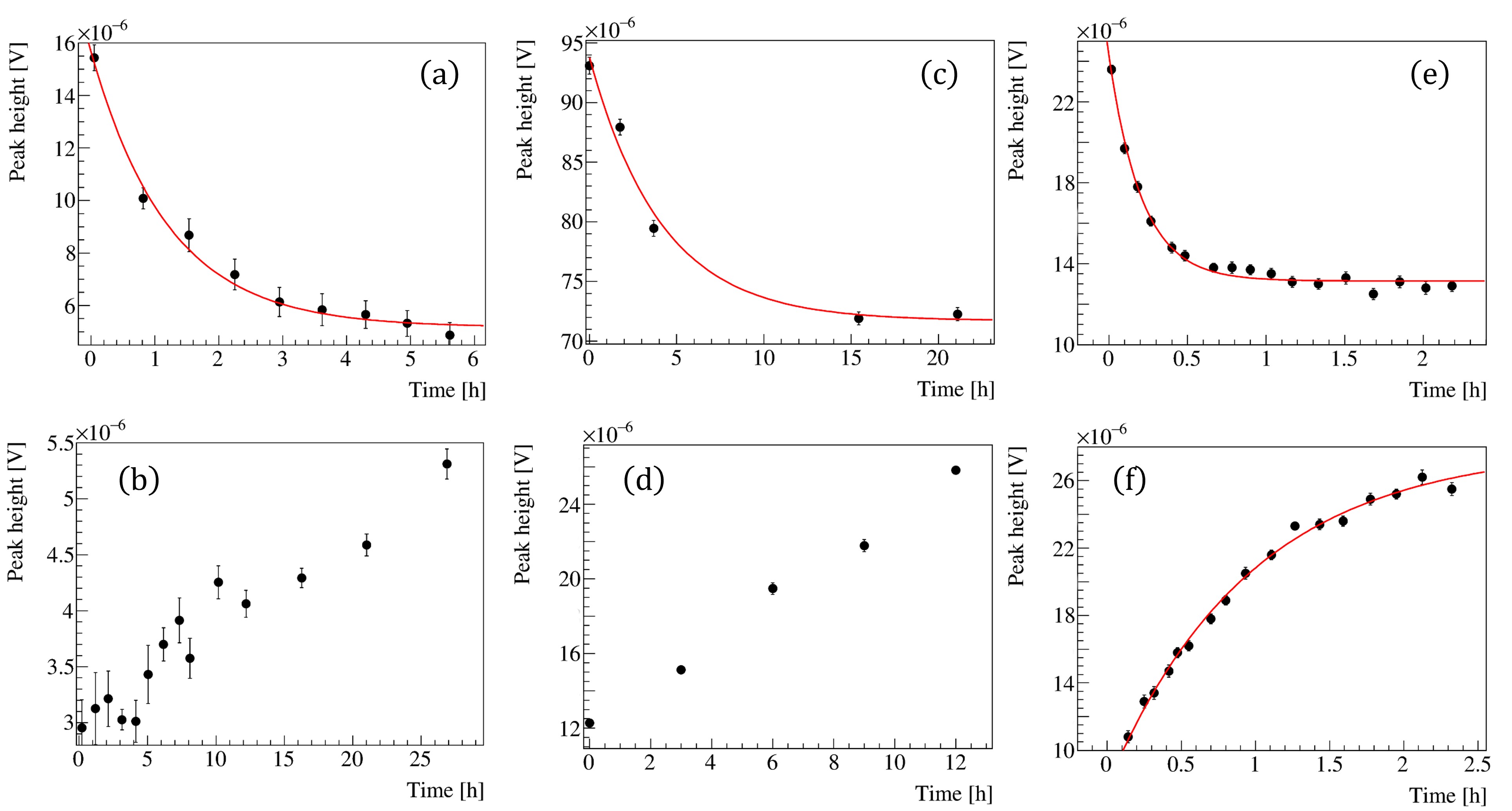}
  \caption{Typical time dependence of the NMR peak heights for 
  (a) $^{139}$La at 0.5 K and 0.5 T, 
  (b) $^{139}$La at 0.1 K and 0.76 T,
  (c) $^{27}$Al at 0.5 K and 2.5 T, 
  (d) $^{27}$Al at 0.1 K and 0.75 T, 
  (e) $^{27}$Al at 1.5 K and 1.0 T, and 
  (f) $^{27}$Al at 1.5 K and 2.0 T.}
  \label{fig:time_evo} 
\end{figure*}

By confirming that the peak widths do not change with time, the peak heights are used for the analysis of the time dependence of the NMR intensities. 
Fig.~\ref{fig:time_evo} (a) and (b) show the typical time dependence of the height of the central peak.
The values of $T_{1}$ can be obtained by fitting Eq.~(\ref{eq:time_evolution}) to the time dependence of the peak heights. 
The results of $T_{1}$ for the central peak under various measurement conditions are summarized in Table~\ref{table:T1_results_La}. 
In the fitting results as well as general long $T_1$ analysis, there are some large values of $\chi^2/$ndf, because the fitting analysis does not contain the systematic uncertainty. The dominant systematic effect is originated from a long term instability of the ground line of the LC circuit in the NMR system. Because of the sharp Q curve, tiny drifting and spiking of the ground level causes huge systematic uncertainty. Generally, it is noted that such systematic effects are practically unavoidable in the long term measurements of the NMR, such as the long $T_1$ measurements.  

\begin{table}[h]
  \centering
  \caption{
  Results of the $T_1$ measurements for $^{139}$La.
  }
  \begin{tabular}{cccc}
  \hline
  \textrm{
    \begin{tabular}{c}
      Temperature \\  magnetic field
    \end{tabular}
  }&
  \textrm{
    \begin{tabular}{c}
      Peak \\  index $i$
    \end{tabular}
  }&
  \textrm{$T_1$\, [h]}&
  \textrm{$\chi^2/$ndf}\\
  \hline
  0.5 K, 2.5 T  & 3 & $1.87 \pm 0.15$ & 52.8/8 \\
  0.5 K, 1.0 T  & 3 & $1.82 \pm 0.26$ & 10.8/11 \\
  0.5 K, 0.5 T  & 3 & $1.21 \pm 0.17$ & 3.4/6 \\
  0.1 K, 0.75 T & 3 & $316 \pm 17$ & 24.2/11\\ \hline
  \end{tabular}
\label{table:T1_results_La}
\end{table}

\subsection{$T_1$ measurements of $^{27}$Al NMR in the DRS2500 refrigerator}

In the same manner as the $^{139}$La, the peak heights can be used as the peak intensities in the analysis of the relaxation time, as in the case of $^{139}$La.
Fig.~\ref{fig:time_evo} (c) shows the height of the central peak in Fig.~\ref{fig:alnmr} (b) as a function of time. 

By fitting Eq.~(\ref{eq:time_evolution}) to the time dependence of the peak heights, we can evaluate $T_{1}$. 
The results for 0.5 K are summarized in Table~\ref{table:T1_results}.
However, the data for 0.1 K could not be analyzed with a simple exponential curve, 
because $T_1$ was extremely large. 
The data analysis for 0.1 K will be discussed in Sec.~\ref{sec:0.1K_analysis}.

\begin{table}[h]
  \caption{
  $T_1$ results for the $^{27}$Al NMR peaks.
  }
  \begin{tabular}{cccc} \hline
  \textrm{
    \begin{tabular}{c}
      Temperature \\  magnetic field
    \end{tabular}
  }&
  \textrm{
    \begin{tabular}{c}
      Peak \\  index $i$
    \end{tabular}
  }&
  \textrm{$T_1$\, [h]}&
  \textrm{$\chi^2/$ndf}\\
  \hline
  0.5 K, 2.5 T 
  & 1 & $3.62 \pm 1.08$ & 4.8/2\\
  & 2 & $1.72 \pm 0.68$ & 5.1/2\\
  & 3 & $5.89 \pm 1.50$ & 7.9/2\\
  & 4 & $5.44 \pm 0.90$ & 2.7/2\\
  & 5 & $4.11 \pm 0.36$ & 13.2/2\\
  & 6 & $7.49 \pm 1.16$ & 1.8/2\\
  & 7 & $7.13 \pm 1.84$ & 0.54/2\\
  & 8 & - \\
  & 9 & - \\
  \hline
  0.5 K, 1.0 T 
  & 1 & $1.65 \pm 0.27$ & 12.2/8\\
  & 2 & $3.47 \pm 0.57$ & 12.3/8\\
  & 3 & $1.87 \pm 0.17$ & 18.4/8\\
  & 4 & $1.87 \pm 0.11$ & 24.8/8\\
  & 5 & $2.15 \pm 0.09$ & 64.9/8\\
  & 6 & $1.94 \pm 0.14$ & 65.4/8\\
  & 7 & $2.05 \pm 0.24$ & 45.1/8\\
  & 8 & $2.74 \pm 0.41$ & 40.1/8\\
  & 9 & $2.52 \pm 0.50$ & 14.8/8\\
  \hline
  0.5 K, 0.5 T 
  & 1 & $2.47 \pm 0.42$ & 8.1/6\\
  & 2 & $2.42 \pm 0.22$ & 47.1/6\\
  & 3 & $1.68 \pm 0.12$ & 53.6/6\\
  & 4 & $1.85 \pm 0.07$ & 109.3/6\\
  & 5 & $2.06 \pm 0.05$ & 132.1/6\\
  & 6 & $1.72 \pm 0.06$ & 151.3/6\\
  & 7 & $1.89 \pm 0.12$ & 27.8/6\\
  & 8 & $2.04 \pm 0.16$ & 17.3/6\\
  & 9 & $1.65 \pm 0.24$ & 15.2/6\\
  \hline
\end{tabular}
\label{table:T1_results}
\end{table}

\begin{table}[h]
  \caption{
  $T_1$ results at 1.5 K for $^{27}$Al.
  }
  \begin{tabular}{cccc} \hline
  \textrm{
    \begin{tabular}{c}
      Temperature \\  magnetic field
    \end{tabular}
  }&
  \textrm{
    \begin{tabular}{c}
      Peak \\  index $i$
    \end{tabular}
  }&
  \textrm{$T_1$\, [h]}&
  \textrm{$\chi^2/$ndf}\\
  \hline
  1.5 K, 1.0 T
  & 5 & $  0.21 \pm 0.01 $ & 17.7/14\\
  1.5 K, 2.0 T
  & 5 & $  0.97 \pm 0.06 $ & 24.5/14\\
  \hline
  \end{tabular}
  \label{table:T1_results_SC}
\end{table}

\subsection{$T_1$ measurements of $^{27}$Al NMR using the SC}

The time dependence of the peak height of the central peak at 1.5 K is shown in 
Fig.~\ref{fig:time_evo} (e) and (f). As in the previous analysis, 
fitting Eq.~(\ref{eq:time_evolution}) to the peak heights yields the relaxation times at 1.5 K. 
The results for 1.0 T and 2.0 T are listed in Table~\ref{table:T1_results_SC}.

\section{Discussion}

\subsection{Estimation of $T_{1}$ at 0.1 K}
\label{sec:0.1K_analysis} 
As shown in the previous section, the analysis using Eq.~(\ref{eq:time_evolution}) is not directly applicable to the $^{27}$Al data at 0.1 K, because the NMR signals are not saturated within the limited experimental period. To make a rough estimation, we use the first approximation of Eq.~(\ref{eq:time_evolution}) with respect to time as follows:

\begin{equation}
  I(t) = I_0 + (I_{\rm eq}-I_0)\Gamma t.
  \label{eq:modified_fitting_function}
\end{equation}
Here, the $I_{\rm eq}$ at 0.1 K is treated as a fixed value and is converted from the $I_{\rm eq}$ at 0.5 K and 1.0 T under the assumption of a Boltzmann distribution. This assumption requires a sufficiently small difference in the NMR sensitivity between the two experimental conditions. The sensitivity of the used system must be checked in this experiment because the corresponding NMR frequency, 16.8 MHz or 16.4 MHz, is different from that at 0.5 K and 1.0 T as presented in Table~\ref{table:dataset}. Consequently, we experimentally verified that the sensitivity for 16.8 MHz, 16.4 MHz, and 11.8 MHz are nearly equivalent. 
Therefore, the $I_{\rm eq}$ in the transition $m-1 \leftrightarrow m$ can be estimated as follows:

\begin{equation}
\begin{split}
    &I_{\rm eq, m-1 \leftrightarrow m}({\rm 0.1 \, K}, B_0) \\
    &\, \approx I_{\rm eq,\,m-1 \leftrightarrow m}({\rm 0.5\,K}, {\rm 1.0 \, T})
    \times \frac{{\rm 0.5\,K}}{{\rm 1.0\,T}} \times \frac{B_0{\rm\,T}}{{\rm 0.1\,K}}  
\end{split}
\end{equation}
The results of the data analysis are summarized in Table~\ref{table:extreme_long_T1_results}. 

\begin{table}[h]
  \caption{
  Estimation of $T_1$ at 0.1 K for $^{27}$Al.
  }
  \begin{tabular}{cccc} \hline
  \textrm{
    \begin{tabular}{c}
      Temperature, \\  magnetic field
    \end{tabular}
  }&
  \textrm{
    \begin{tabular}{c}
      Peak \\  index $i$
    \end{tabular}
  }&
  \textrm{$T_1$\, [h]}&
  \textrm{$\chi^2/$ndf}\\
  \hline
  0.1 K, 0.76 T 
  & 1 & $  53 \pm   3 $ & 2.7/3 \\
  & 2 & $  40 \pm   2 $ & 5.9/3 \\
  & 3 & $  78 \pm   6 $ & 5.8/3 \\
  & 4 & $  71 \pm   3 $ & 6.1/3 \\
  & 5 & $  54 \pm   1 $ & 10.1/3 \\
  & 6 & $  81 \pm   3 $ & 11.0/3 \\
  & 7 & $  86 \pm   7 $ & 7.3/3 \\
  & 8 & $  45 \pm   2 $ & 6.6/3 \\
  & 9 & $  49 \pm   3 $ & 0.3/3\\
  \hline
  0.1 K, 1.6 T 
  & 1 & - \\
  & 2 & $ 546 \pm  48 $ & 4.1/2 \\
  & 3 & $ 293 \pm  12 $ & 5.9/2 \\
  & 4 & $ 339 \pm  10 $ & 4.5/2 \\
  & 5 & $ 397 \pm   8 $ & 13.8/2 \\
  & 6 & $ 363 \pm  11 $ & 19.3/2 \\
  & 7 & $ 388 \pm  11 $ & 2.6/2 \\
  & 8 & $ 356 \pm  18 $ & 1.7/2 \\
  & 9 & $ 235 \pm   8 $ & 19.7/2 \\
  
  \hline
  \end{tabular}
  \label{table:extreme_long_T1_results}
\end{table}

\subsection{Comparison of $T_{1}$ between the $^{139}$La and $^{27}$Al nuclei}
Table~\ref{table:T1_results_La_and_Al} compares the $T_{1}$ in the $-1/2 \leftrightarrow 1/2$ transitions of the $^{139}$La and $^{27}$Al under the same conditions. All $T_{1}$ values of both the nuclei at 0.5 K are roughly comparable, although the $T_{1}$ of $^{139}$La tends to be slightly shorter than that of $^{27}$Al. However, the $T_{1}$ of $^{139}$La at 0.1 K is approximately six times that of $^{27}$Al. The relation of the $T_{1}$ at 0.1 K is substantially different from that at 0.5 K.

According to previous DNP studies with Nd-doped LaAlO$_{3}$ crystals~\cite{Hautle, Maekawa1995}, the nuclear Zeeman systems of both $^{27}$Al and $^{139}$La strongly couple to the electronic spin-spin reservoir (SSR)~\cite{order_and_disorder}, and their thermal contact contributes to the large enhancement of the nuclear polarization in the LaAlO$_{3}$ crystals. In such a system, the nuclear relaxation is likely to occur through the SSR. In Ref.~\cite{Maekawa1995}, coupling strengths of a few minutes were reported for some central peaks for $^{139}$La. If the SSR dominantly contributes to the nuclear relaxation, the $T_{1}$ values of both nuclear species can be expected to be almost identical, because the two Zeeman systems and the SSR can be considered a single reservoir with a common spin temperature. 
The results at 0.5 K are consistent with the nuclear relaxation process via the SSR. If the SSR is also the dominant process at 1.5 K, the results for $^{27}$Al can be regarded as the relaxation of $^{139}$La at 1.5 K, and the $T_{1}$ at 2.0 T is 1 h. 

A possible explanation for the large difference in $T_1$ between $^{139}$La and $^{27}$Al at 0.1 K is the strong temperature dependence at approximately 0.1 K. In the relaxation process through the SSR, the relaxation rate is proportional to the factor $1-P_0^2$, where $P_0$ is the polarization of the paramagnetic impurities~\cite{order_and_disorder} in thermal equilibrium and is expressed as 

\begin{equation}
P_0 = \tanh \left( \frac{g_{\rm Nd^{3+}} \mu_B B_0}{2k_BT} \right).
\label{eq:Nd_polarization}
\end{equation}
Here, $g_{\rm Nd^{3+}}$ is the effective g factor of Nd$^{3+}$ in the LaAlO$_3$ crystal, and its value in the primary domain is $2.12\pm0.01$~\cite{Takahashi1993}. The simple calculation of $1-P_0^2$ shows that the relaxation time at 0.11 K is approximately three times shorter than that at 0.10 K under a magnetic field of 0.75 T. Therefore, the temperature uncertainty of 5\% is not negligible under this condition. 
The sample was removed from Cold-Finger2 after the NMR measurement of $^{27}$Al and mounted again for that of $^{139}$La.
As the temperature of 0.1 K was always monitored using a thermometer attached to the mixing chamber, it is likely that the sample temperature is higher than 0.1 K. 

The $T_1$ difference between the primary and secondary domains at 0.1 K and above 0.75 T in Table~\ref{table:extreme_long_T1_results} can be explained by the anisotropy of $g_{\rm Nd^{3+}}$ ($g_{\parallel}=2.12, g_{\perp}=2.68$) because, in the secondary domain, the symmetry axis is not parallel to the magnetic field and the effective $g$ factor is larger than that in the primary domain~\cite{Takahashi1993}.

\begin{table}[h]
  \centering
  \caption{
  Comparison of the $T_1$ for the transition $m_I:-1/2 \leftrightarrow 1/2$ of $^{139}$La and $^{27}$Al in the Nd-doped sample under various conditions.}
  \begin{tabular}{ccc} \hline
    \textrm{
    \begin{tabular}{c}
      Temperature \\  magnetic field
    \end{tabular}
  }&
  \textrm{$T_{1}$ of $^{139}$La\,[h]}
  &\textrm{$T_{1}$ of $^{27}$Al\,[h]}\\
  \hline
  0.5 K, 0.5 T & $ 1.21 \pm 0.17 $ & $ 2.06 \pm 0.05 $ \\
  0.5 K, 1.0 T & $ 1.82 \pm 0.26 $ & $ 2.15 \pm 0.09 $ \\
  0.5 K, 2.5 T & $ 1.87 \pm 0.15 $ & $ 4.11 \pm 0.36 $ \\
  0.1 K, 0.75 T& $ 316 \pm 17 $ & - \\
  0.1 K, 0.76 T&  -  & $  54 \pm 1 $ \\
  \hline
  \end{tabular}
  \label{table:T1_results_La_and_Al}
\end{table}

\subsection{Feasibility of the T-violation searches with the Nd-doped LaAlO$_3$ crystal}
When we use the polarized $^{139}$La target in the search for the T-violation, it is necessary to decrease the external magnetic field for canceling out the neutron spin rotations with the pseudomagnetic effect, 
which is caused by the interaction between an incident neutron and a polarized nucleus~\cite{order_and_disorder}. It possibly 
causes significant systematic uncertainty and a reduction in the sensitivity for the T-violation search. Previous studies have demonstrated that the magnitude of the external magnetic field for canceling pseudo magnetic rotation is approximately 0.23 T when using a 100 \% polarized LaAlO$_{3}$ target~\cite{Takahashi1994}. 
Considering the past achievement of approximately 50 \% polarization, it can be inferred that the magnetic field in the target operation is approximately 0.1 T. Considering the cryogenic system in a high-intensity neutron beamline, a realistic condition for the temperature is approximately 0.1 K. The $T_{1}$ of $^{139}$La under the above target condition is rather significant, but most NMR apparatuses including our system are not generally adaptable for such a low magnetic field. Thus, we roughly estimated $T_{1}$ at 0.1 K and 0.1 T from our experimental results.

\begin{figure}[h]
  \centering
  \includegraphics[width=8.5cm]{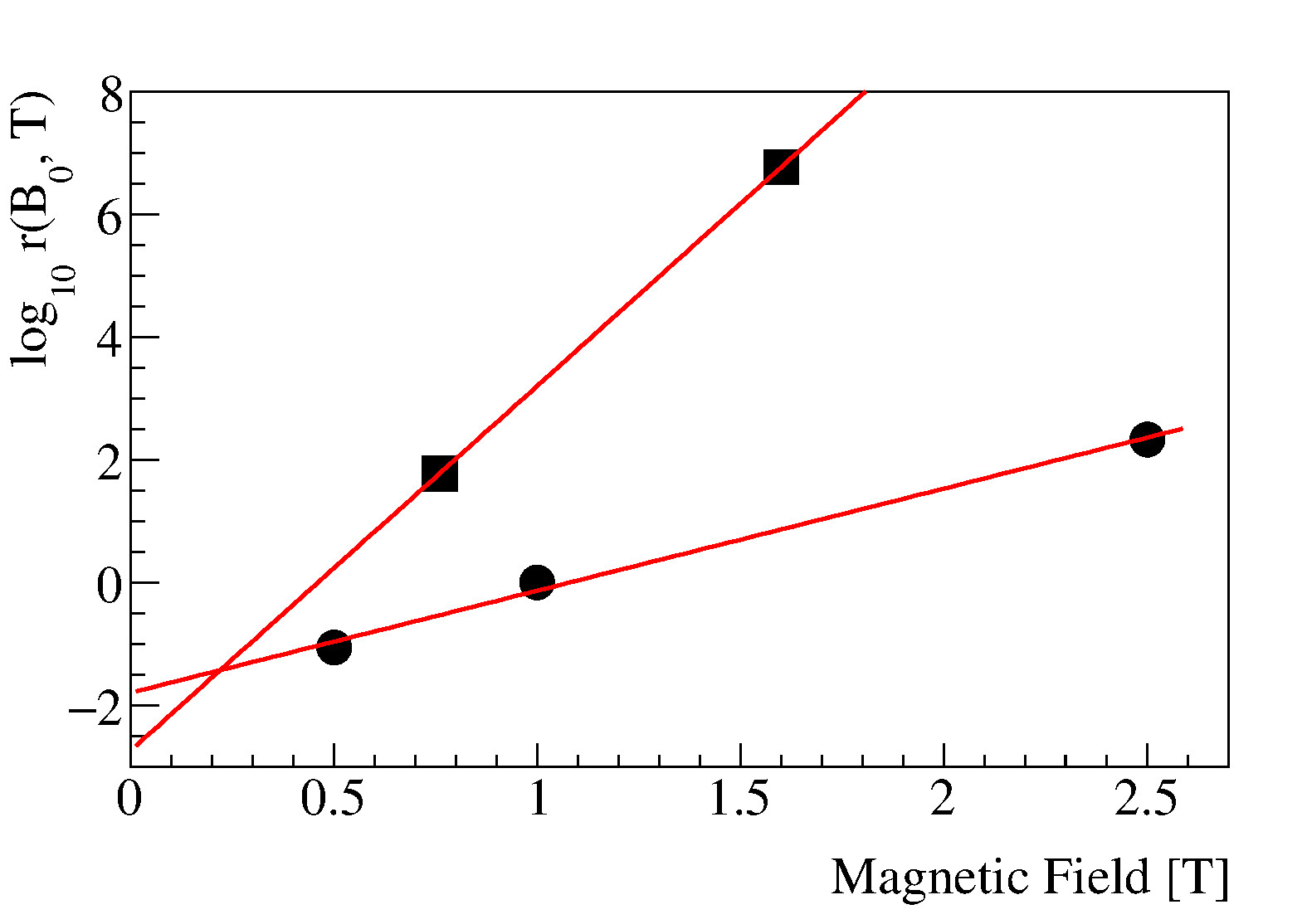}
  \caption{$T_{\rm 1ss}$ estimated from the $^{27}$Al data using the simple relaxation model described by Eq.~(\ref{eq:relaxation_model}). The circles and squares are the estimated $\log_{10} r$ at 0.5 K and 0.1 K, respectively.}
  \label{fig:T1ss} 
\end{figure}

We applied a simple relaxation model via the SSR~\cite{order_and_disorder}, which is described as, 

\begin{equation}
  \frac{1}{T_1} \propto C^2 \frac{1}{T_{\rm 1ss} } \left( \frac{1}{B_0} \right)^2 (1-P_0^2),\\  
  \label{eq:relaxation_model}
\end{equation}
where $T_{\rm 1ss}$, $C$, and $P_0$ are the relaxation time of the SSR, the concentration of Nd$^{3+}$, and the polarization of Nd$^{3+}$ given by Eq.~(\ref{eq:Nd_polarization}), respectively. Using Eq.~(\ref{eq:relaxation_model}), the $T_{1}$ results for $^{27}$Al facilitate the deduction of the ratios of $T_{\rm 1ss}$ to that under a specific condition for various conditions:

\begin{equation}
\begin{split}
  &r(B_0, T) \coloneqq \frac{T_{\rm 1ss}({\rm 1.0 \, T, 0.5 \, K})}{T_{\rm 1ss}(B_0, T)} = \\
  &\frac{T_1({\rm 1.0 \, T, 0.5 \, K})}{T_1(B_0, T)} \frac{B_0^2}{({\rm 1.0\, T})^2} \frac{1-P_0({\rm 1.0 \, T, 0.5 \, K})^2}{1-P_0(B_0, T)^2}.
\end{split}
\end{equation}
Fig.~\ref{fig:T1ss} shows $\log_{10} r(B_0, T)$ as a function of the magnetic field. 

Because the coupling strength of the SSR to the lattice system has not been sufficiently studied in terms of the field dependence so far, we simply extrapolate the $T_{\rm 1ss}$ from Fig.~\ref{fig:T1ss} here. Consequently, the field dependence of the relaxation rate $1/T_{\rm 1ss}$ is approximately identified as an exponential decay with respect to the magnetic field. As Fig.~\ref{fig:T1ss} shows that the $T_{\rm 1ss}$ at 0.1 K and 0.1 T is approximately 100 times that at 0.5 K and 1.0 T, we can obtain the $T_{\rm 1}$ of $^{139}$La under the target condition as

\begin{eqnarray}
  T_1 \approx T_1({\rm 1.0\,T,0.5\,K,\,La}) \left( \frac{0.1\,{\rm T}}{1.0\,{\rm T}} \right)^2 \times 0.01 \nonumber \\ 
  \times \frac{ 1 - P_0({\rm 1.0\,T,0.5\,K})^2 }{1 - P_0({\rm 0.1\,T,0.1\,K})^2}. 
  \label{eq:relaxation_estimation}  
\end{eqnarray}
$T_{1}$ is obtained as approximately 1 h. However, this analysis is based on the results of $^{27}$Al at 0.1 K. 
Assuming that the large difference of $T_1$ at 0.1 K between $^{139}$La and $^{27}$Al in Table~\ref{table:T1_results_La_and_Al} is caused by a slight difference in the sample temperature, it is expected that the $T_1$ of $^{27}$Al was underestimated. 
Therefore, we can expect that $T_{1}\geq 1$ h under the target condition.

There is still much room for improving the nuclear relaxation in terms of the following three points. The first is to decrease $C$ in Eq.~(\ref{eq:relaxation_model}) because the nuclear relaxation through the SSR depends on the square of $C$. If $C$ is reduced by a factor of three compared to the present value, for example, $T_{1}$ can be expected to become an order of a large magnitude. The second is the preparation of a crystal with high quality that is almost free from crystal defects such as oxygen vacancies. Such defects act as paramagnetic centers, which accelerate the nuclear relaxation. The third is to use a powerful dilution refrigerator to reduce the temperature to 0.05 K.

\section{Summary}

To evaluate the possibility of using a LaAlO$_{3}$ crystal doped with 0.03 mol\% Nd ions as the polarized La target, we conducted the first experiment to systematically measure the nuclear spin-lattice relaxation time of $^{139}$La and $^{27}$Al in the crystal at temperatures of 0.1, 0.5, and 1.5 K under various magnetic fields up to 2.5 T. 
Additionally, we estimated the relaxation time at 0.1 K under 0.1 T, which is a suitable condition for beam experiments.

The relaxation time of $^{139}$La at 0.5 K was 1.21$\pm$0.17 h, 1.82$\pm$0.26 h, and 1.87$\pm$0.15 h for 0.5, 1.0, and 2.5 T, respectively, and that at 0.1 K was 316$\pm$17 h for 0.75 T. These results at 0.5 K are comparable to those of $^{27}$Al and consistent with the nuclear relaxation, which is dominant through the SSR. The significant difference at 0.1 K can be explained from the temperature uncertainty of 5 \% under the assumption of the SSR. We simply extrapolated the relaxation time of the SSR with the results of $^{27}$Al. Consequently, we estimated that the $T_{1}$ of $^{139}$La is greater than 1 h at 0.1 K and 0.1 T.

There is still much room for improving the value, particularly by reducing the doping amount of the Nd ions, using a crystal with a low defect density, and cooling down to 0.05 K using powerful cryogenics. 
An improvement by two orders of magnitude will enable the use of the Nd-doped LaAlO$_{3}$ crystal in T-violation experiments. 
To improve the present nuclear relaxation, crystal growth and powerful cryogenics are expected to be key factors. We will continue the studies on the crystal growth of Nd-doped LaAlO$_3$ crystals using the floating-zone method~\cite{Ishizaki}. This method is expected to avoid undesirable contamination by impurities originating from the use of the crucible. 

\section*{Acknowledgement}
The present study was conducted at the Research Center for Nuclear Physics (RCNP), 
Osaka University with the approval of the project proposal 
"Development of polarized target for new physics search via T-violation." This work was partly supported by the RCNP Collaboration Research Network
program as the project number COREnet-026. The studies on the target material were supported by the Tohoku University 
IMR cooperative program (proposal No. 18G0034, 19K0081, and 19G0037). 
We are grateful to the staff of the Low Temperature Center of 
Osaka University for supplying us with the required liquid helium. 
We thank Dr. Patrick Hautle (PSI) and Prof. Taku Matsushita (Nagoya Univ.) for their fruitful discussions. 
The present work was supported in part by the Ministry of Education, 
Science, Sports, and Culture of Japan.

\end{document}